\documentclass[american, aps,prl, twocolumn, showpacs,reprint]{revtex4} 
\usepackage[unicode=true,pdfusetitle, bookmarks=true,bookmarksnumbered=false,bookmarksopen=false, breaklinks=false,pdfborder={0 0 0},backref=false,colorlinks=false] {hyperref}
\hypersetup{ colorlinks,linkcolor=myurlcolor,citecolor=myurlcolor,urlcolor=myurlcolor}
\usepackage{braket,colortbl,cleveref,amsthm,amsmath,amssymb,txfonts}
\definecolor{myurlcolor}{rgb}{0.0,0.0,0.7}
\usepackage{graphics,graphicx}
\usepackage{color}
\usepackage{graphicx}
\usepackage{graphicx}  
\usepackage{booktabs}
\usepackage[font=small,labelfont=bf,
   justification= centerlast,
   format=hang]{caption}
\usepackage{subfigure}
\usepackage{tabularx}
\usepackage{amsmath}
\DeclareMathOperator{\tr}{Tr}
\usepackage{graphicx}
\usepackage[utf8x]{inputenc}
\usepackage{color}
\usepackage{amsmath}
\usepackage{braket}
\usepackage{latexsym}
\usepackage{amssymb}
\usepackage{amsthm}
\usepackage{bm}
\usepackage{graphics,epstopdf}
\usepackage{color}\usepackage{amsmath}
\usepackage{enumitem}
\usepackage{fmtcount}
\usepackage{booktabs}

\usepackage{epsfig}

\theoremstyle{plain}

\def\bea{\begin{eqnarray}}
\def\eea{\end{eqnarray}}
\def\ba{\begin{array}}
\def\ea{\end{array}}

\def\ket{\rangle}
\def\bra{\langle}
\def\beq{\begin{equation}}
\def\eeq{\end{equation}}

 \newtheorem{theorem}{Theorem}

\begin{document}

\title{Stronger classes of  sum uncertainty and reverse uncertainy relations}

\author{Chiranjib Mukhopadhyay}
\author{Arun Kumar Pati}
\address{Quantum Information and Computation Group, Harish Chandra Research Institute, Homi Bhabha National Institute, Allahabad 211019, India}

\begin{abstract} 
\noindent 
 
Uncertainty relations are old, yet potentially rewarding to explore. By introducing a quantity called the \emph{uncertainty matrix}, we provide a link between purity and observable incompatibility, and  derive several stronger uncertainty relations in both forward and reverse directions for arbitrary quantum states, i.e., mixed as well as pure, and arbitrary incompatible quantum observables, none of which suffer from the problem of triviality. Besides the tightness, the interpretations of terms in these uncertainty relations may be of independent interest. We provide the possible generalization of stronger uncertainty relations to sum of variances of more than two observables. We also demonstrate  applications of techniques used here to, firstly, obtain a simple reverse quantum speed limit for quantum states undergoing Markovian dynamical evolution, and secondly, to provide a lower bound for fidelity between two quantum states.

 
\end{abstract}

\maketitle

\emph{Introduction-} Uncertainty relations are at the heart of quantum theory. Almost a century after the discovery of quantum theory, one may thus be surprised by the recent resurgence in research on preparation uncertainty relations \citep{mpur, maziero, urchin1, urchin2, urchin3, branciardur, mdur, debasis, bialynicki, maassen, wehner_review, huang, schwonnek}, as well as on the exact interpretation and formulation of Heisenberg's uncertainty principle \citep{ozawa,experimental, pekka, ozawa_mixed, branciard, namrata, busch_review}. The problem of triviality of the uncertainty relation was recently addressed in \cite{mpur}, but there is still ample scope for  devising stronger uncertainty relations. In addition to the fundamental importance of these developments, they allow us to come up with applications, such as quantum speed limits \citep{tamm, anandan, del_campo,  seventh, deffner_review} or, detection of nonclassical correlations \citep{hofmann, guehne, cavalcanti} among many others. Thus, the discovery of any new uncertainty relation may have important potential implications for quantum technology.

In this letter, we outline several new uncertainty relations. Firstly, we extend the formalism of Ref. \cite{mpur} to obtain non-trivial lower bounds for sum of variance-based uncertainties for all states, not just pure states or \emph{some} mixed states. Subsequently, we formulate yet another sum uncertainty relation, which is optimization-free, and nicely separates out the quantum and the classical contributions to uncertainty. Next, we obtain a very strong uncertainty relation for arbitrary dimensional systems, perhaps one of the strongest variance based state-dependent uncertainty relations in literature.  This is followed by an extension of these ideas to the case of sum uncertainty relations for arbitrary number of observables. In addition to the uncertainty relations, we also formulate a family of reverse uncertainty relations, and mention two applications of these relations. The first is to formulate a lower bound for speed of quantum evolution, as opposed to the upper bound which has been the subject of considerable interest  throughout the history of quantum theory \citep{deffner_review}, for an initially mixed qubit system undergoing Markovian evolution. The second aims at figuring out a lower bound on the fidelity between two arbitrary states.


\emph{Uncertainty matrix -} Let $A$ and $B$ be two observables. Let us introduce the following Hermitian operators $C = A - \bra A \ket, D = B - \bra B \ket$. We term the following matrix, $K$ as the \emph{uncertainty matrix}.
\begin{equation}
K = (C \pm iD)(C \mp iD) 
\label{urmat}
\end{equation}
 Since $K$ is expressed in the form $X^{\dagger}X$, where $X = C \pm iD$, it is guaranteed to be Hermitian and positive semi-definite. For qubit systems, it has been shown in the supplementary material, that the purity of the density matrix thus constructible from the uncertainty matrix via normalization, captures the intrinsic incompatibility of the two operators. This provides a clue to link the resource theory of purity \citep{purity} with quantum uncertainty.  Let us note the following identity for the uncertainty matrix, derivable from the parallelogram law \citep{mpur},  which holds for any pure state $|\psi\rangle$. 
\begin{equation}
 || (C \mp iD) |\psi \rangle ||^2  = \Delta A^{2} + \Delta B^{2} \mp  i \langle [A,B] \ket | 
\label{urequality}
\end{equation}


\emph{Stronger uncertainty relation for arbitrary mixed states-} The problem of possible triviality of the variance based Robertson-type sum uncertainty relation even in the case of non-commuting observables was recently resolved in Ref. \citep{mpur} for pure states, but still left the case of arbitrary mixed states open. Let us now formulate an analogus sum uncertainty relation valid for arbitrary mixed states utilizing the vectorization operation \citep{supp}, which entails constructing a vector $|\rho\rangle$ from a matrix $\rho$ by stacking columns of $\rho$ on top of each other. The supplementary material contains relevant properties of vectorization, as well as an intriguing result linking entanglement, coherence, purity, and imaginarity of qubit systems via the vectorization technique.

As before, let $A - \langle A \rangle = C$, and $B - \langle B \rangle = D$. Now, $\Delta A = \sqrt{\text{Tr} (\rho C^2)} = \sqrt{\text{Tr} (\sqrt{\rho} C^2 \sqrt{\rho})} = \sqrt{\langle \sqrt{\rho} | \mathbb{I} \otimes C^2 | \sqrt{\rho} \rangle}$. Thus, $\Delta A =  || (\mathbb{I} \otimes C ) |\sqrt{\rho} \rangle ||$. Similarly one can show that $\Delta B =   || (\mathbb{I} \otimes iD ) |\sqrt{\rho} \rangle||$. The vectors here can be normalized to unity and, thus be considered as pure states in a $d^2$-dimensional Hilbert space, if the original system was $d$-dimensional. Since square root of a matrix is non-unique, we can simply choose that square root which is Hermitian. A possible prescription for constructing such a matrix is to first diagonalize $\rho$ via an unitary $U$, then take the positive square roots of the population elements, then finally apply the reverse unitary $U^{\dagger}$.  From \eqref{vecinnerpdt}, it follows that the vectorization of this matrix is normalized, and thus, a quantum state vector.  Now, we are in the position to prove the following mixed state generalization to the first inequality in \citep{mpur}.

\begin{theorem}
If $|k^{\perp}\rangle$ is a (normalized) state vector perpendicular to the vectorization $|k\rangle$, the following variance based sum uncertainty relation holds- 
\begin{equation}
\Delta A^2 + \Delta B^2 \geq \pm i \langle [A,B] \rangle + |\bra\sqrt{\rho}|  \mathbb{I} \otimes (A \pm iB)  |\sqrt{\rho}^{\perp} \rangle |^2.
\end{equation}
\label{mixed_ur_thm}
\end{theorem}
\emph{Proof-} Upon simplification, like before, $|| (C \pm iD |\sqrt{\rho}) \rangle ||^2 = \Delta A^2 + \Delta B^2 \mp i \langle [A,B] \rangle$. Now, by applying the Cauchy-Schwarz inequality on two vectors $(C\pm iD)|\rho\rangle$ and $|\rho^{\perp}\ket$, the latter being a (normalized) quantum state vector orthogonal to $|\sqrt{\rho}\rangle$, and using the parallelogram law,we have $\Delta A^2 + \Delta B^2 \mp i \langle [A,B] \rangle   \geq  |\langle \sqrt{\rho}| A\pm iB - (\langle A \rangle + \langle iB \rangle)| \sqrt{\rho}^{\perp}\ket|^2 = |\langle\sqrt{\rho}|  \mathbb{I} \otimes (A \pm iB) |\sqrt{\rho^{\perp}} \rangle |^2$. This completes the proof.

 \qed

An important point to note here is that, unlike the vector $|\sqrt{\rho}\ket$, the orthogonal vector $|\sqrt{\rho}^{\perp}\ket$ need not come via vectorization from a Hermitian matrix.  Therefore, optimizing over all such orthogonal vectors turns out to be much tighter than if we restrict to all orthogonal vectors derived from the vectorization of a $d-$dimensional Hermitian matrix. 

\noindent \emph{An optimization-free uncertainty relation-} The uncertainty relations formulated above all depend on optimization over an infinite set of states. While this lets one considerably tighten the corresponding inequalities, it is perhaps desirable to search for a sum uncertainty relation which retains the advantage of non-triviality, but, in addition, is also free from any optimization. Thus, we now formulate the following optimization-free variance based sum uncertainty relation.

\begin{theorem}
For arbitrary mixed state $\rho$ and observables $A$ and $B$, if the corresponding uncertainty matrix is denoted by $K$, then the following optimization-free variance based sum uncertainty relation holds -
\begin{equation}
\Delta A^{2} + \Delta B^{2} \geq |\langle [A,B] \rangle | + S(\rho) - \ln \tr \left( e^{-K} \right), \label{peierls}
\end{equation}
where $S(\rho)$ is the von Neumann entropy of the quantum state $\rho$.
\end{theorem}

\noindent \emph{Proof- } For any mixed state $\rho$,  \eqref{urequality} may be written as \[\tr (\rho K) = \Delta A^2 + \Delta B^2 \mp i \langle [A,B]\rangle. \]  Now let us recall the Peierls-Bogoliubov inequality \citep{bhatia} for the positive observable $K$ and density matrix $\rho$ \beq \tr(\rho K) + \tr (\rho \ln \rho) \geq - \ln \tr e^{-K}. \eeq Combining the above results immediately lead to the proof. \qed\\
This inequality is saturated provided $\rho = e^{-K}$. Since the uncertainty matrix $K$ implicitly depends on the choice of density operator $\rho$, if one fixes the observables, this condition leads to a transcendental equation for the density matrix $\rho$. In the supplementary material \citep{supp}, we comment on the solution of this equality condition in mixed qubit systems. The uncertainty relation above can also be extended in terms of generalized entropies utilizing a recent extension of the Peierls-Bogoliubov inequality for deformed exponentials \citep{hansen}. Let us now comment on the terms that appear in the RHS of the uncertainty relation above.

The uncertainty, quantified by variances, for a mixed quantum state can be thought as originating from two different sources of randomness. One is the intrinsic randomness native to the quantum formalism, the other is the randomness introduced by the classical statistical mixture in forming the mixed state. Thus, when we consider the uncertainty relations as a pillar of quantum mechanics, it is desirable to separate out these two contributions explicitly. This was first attempted by Luo, who introduced the Wigner-Yanase skew information as a quantifier of the `quantum' part of the uncertainty \citep{luo}.  The uncertainty relation \eqref{peierls} above offers an alternative view of approaching this problem. The first term in RHS is the commutator and can be thought of as the intrinsic quantum contribution. The second term is the entropy of the state, which arises from randomness entirely due to the classical mixing. The final term depends only on the uncertainty matrix, which is free from explicit state dependence if the classical averages, i.e., expectation values, of operators $A$ and $B$ are already specified. Thus, the uncertainty lower bound in \eqref{peierls} can be thought of as sum of three distinct contributions. The first being that due to inherent non-commutativity of quantum mechanics, the second entirely due to the classical randomness introduced via mixing, and the third representing, in some sense, a  state-independent contribution to the uncertainty. 

\begin{figure}
 \includegraphics[width=0.45\textwidth, keepaspectratio]{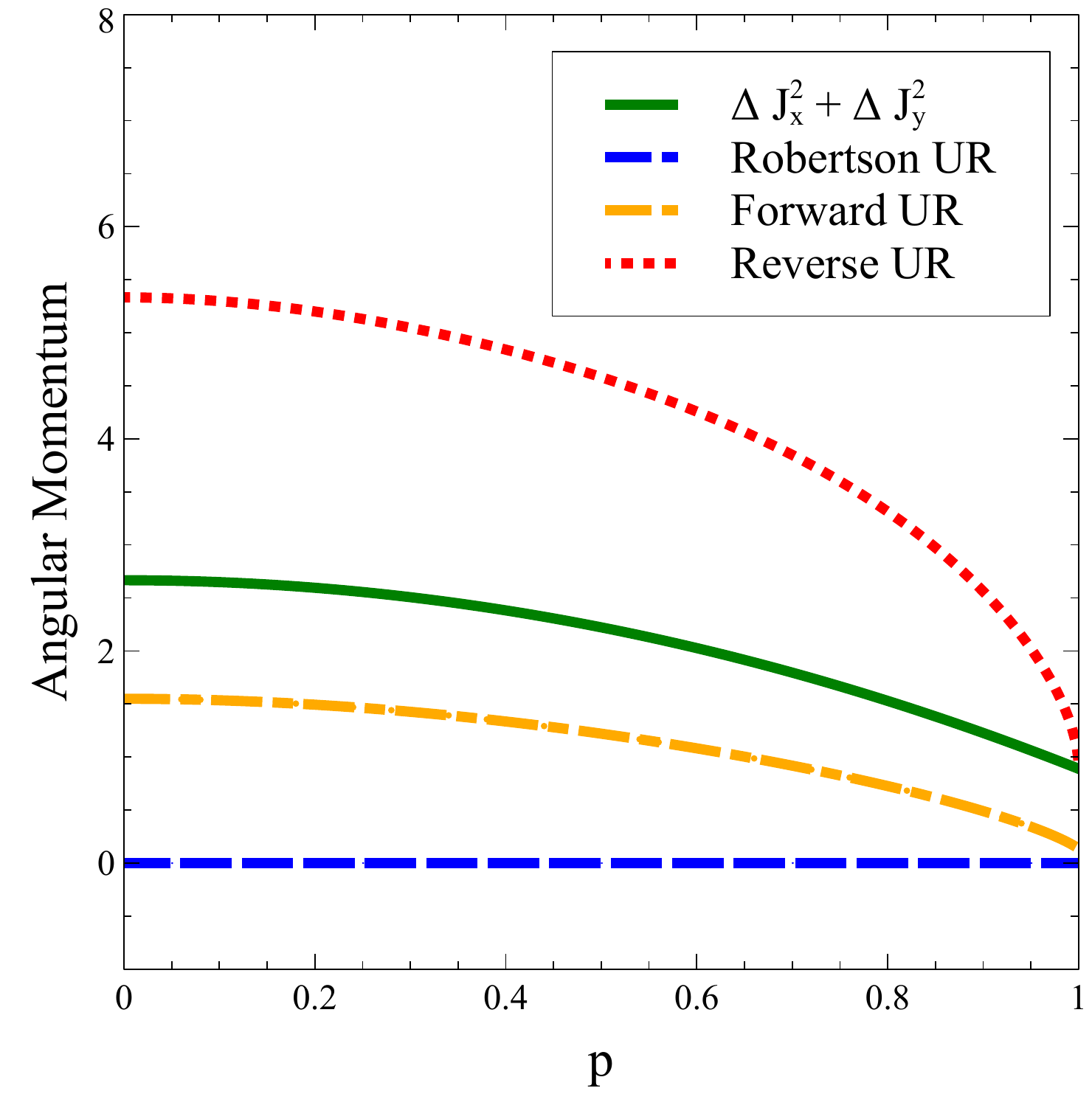}       
\caption{(Color online) Demonstration of the optimization-free sum uncertainty relation \eqref{peierls} (golden dash-dotted line) and the first reverse uncertainty relation (red dotted line)  \eqref{reverse_eqn} with respect to the sum of variances (green line) and the lower bound in Robertson's uncertainty relation (blue dashed line) for the family of qutrit states $\rho = p |\psi \ket \bra \psi| + \frac{1-p}{3} \mathbb{I}$, where $|\psi\ket = \frac{1}{\sqrt{3}} \left(|0\ket + |1\ket + |2\ket  \right)$. The observables are $J_x$ and $J_y$ respectively. }
\label{pb_reverse_fig}
\end{figure}

\noindent \emph{Even stronger uncertainty relation for arbitrary pure  states-} Having dealt with the triviality problem of Heisenberg uncertainty relation for arbitrary quantum states, let us now present a further tightening of the stronger uncertainty relation originally enunciated in \citep{mpur}, which holds for pure states.
\begin{theorem}
For arbitrary pure state $|\psi\ket$  and  observables $A$ and $B$, if the corresponding uncertainty matrix is denoted by $K$, then the following variance based sum uncertainty relation holds for all pure states $|\phi\ket$

\begin{equation}
\Delta A^{2} + \Delta B^{2} \geq |\langle [A,B] \rangle | + \frac{1}{\langle K \rangle_{\phi}} \frac{|\langle \phi | K | \psi \rangle |^2}{\cos \left( 2 \cot^{-1} (\alpha \cot \frac{\theta}{2}) \right)},
\label{thm1}
\end{equation}
where $\theta$ is the inner product angle between $|\psi\ket$ and $|\phi\ket$, and $\alpha = \sqrt{\lambda_{\max} / \lambda_{\min}}$, where $\lambda_{\max}$ and $\lambda_{\min}$ are, respectively, the maximum and minimum eigenvalues of the observable $K$. 
\end{theorem}
\emph{Proof- } Let us recall the Bauer-Householder inequality for arbitrary vectors $\vec{x}, \vec{y}$ making an inner product angle $\xi$ between themselves and an invertible matrix $A$ of the commensurate dimension \beq |(A\vec{y})^{\dagger} (A \vec{x}) | \leq ||A \vec{x} || \ ||A \vec{y}|| \cos \Upsilon, \eeq where $\Upsilon = 2 \cot^{-1}\left[\frac{\lambda_{\text{max}} (A)}{\lambda_{\text{min}} (A)} \cot (\xi /2 )  \right]$. Applying this inequality with the uncertainty matrix $K$ chosen as the matrix $A$ above, along with the uncertainty equality \eqref{urequality} leads to the uncertainty relation stated above. \qed \\

The conditions for equality in the Bauer-Householder inequality \citep{liu} for vectors $ \lbrace |\psi\ket, |\phi\ket \rbrace$ and positive semi definite operator $K$ with eigenvectors $|\max\ket$ and $|\min\ket$ corresponding to its maximum and minimum eigenvalues respectively  are any of the following 

\begin{enumerate}
\item $|\psi\ket$ and $|\phi\ket$ are equal upto some phase.

\item The states $|\psi\ket = \frac{1}{\sqrt{2}} \left( \xi \sqrt{1 + |m |} |\lambda_{\max}\ket - \eta \sqrt{1 - |m|}|\lambda_{\min}\ket \right)$, and  $|\phi\ket = \frac{\epsilon}{\sqrt{2}} \left( \xi \sqrt{1 + |m|} |\lambda_{\max}\ket +  \eta \sqrt{1 - |m|} |\lambda_{\min}\ket \right)$ , where $|\xi| = |\eta| = |\epsilon| = 1$, $m = \cos \theta$, and $|\lambda_{\max}\ket$, $|\lambda_{\min}\ket$ are eigenvectors of the uncertainty matrix $K$ corresponding to the maximum and minimum eigenvalues of $K$ respectively.
\end{enumerate}

Thus, this inequality is tight if one chooses $|\psi\ket = |\phi\ket$. However, this is experimentally not very economical since so choosing $|\phi\ket$  requires one to perform the complete state tomography on the system state anyway. Selecting an orthogonal state to the original state is easier. Thus, this inequality can be tight even if $|\psi\ket$ is orthogonal to $|\phi\ket$, provided the second condition above is met. The key improvement over Ref. \citep{mpur} lies in the fact that if we only have access to arbitrary states, not necessarily orthogonal to the system state, then it is possible to obtain a non-trivial tightening of the Robertson UR vide Eq. \eqref{thm1}, which was not possible via Ref. \citep{mpur}. In particular, for qubit systems, the \emph{maximization over all states orthogonal to $|\psi\ket$} prescription doesn't work simply because once a state is specified  in two dimensions, the orthogonal pure state is automatically uniquely determined. However, Eq. \eqref{thm1}, written in terms of states whose Bloch vectors make an arbitrary angle with the Bloch vector of the system qubit, still works. The generalization of Eq. \eqref{thm1} for arbitrary mixed states may be similarly shown via the vectorization procedure in a similar way to the proof of \eqref{mixed_ur_thm}.

\emph{Extension for arbitrary number of observables -} In this work, we have considered the sum uncertainty relations for two observables so far. However, we are often interested in sum of variances for more number of observables, e.g., for entanglement or nonlocality detection \citep{hofmann, guehne, cavalcanti}. Thus, it is imperative that we try to extend our formalism for sum of variances of arbitrary number of observables $A_1, A_2,..., A_n$. A geometrical result, which generalizes the parallelogram law for $n$-vectors \citep{douglas} helps us formulate stronger uncertainty relations in these cases. As the simplest generalization, the case of three-observable sum uncertainty relations has been studied in the supplementary material \citep{supp}. These results allow us to derive the many-observable versions of the uncertainty relations derived in \citep{mpur} as well as the present work\citep{supp}. We hope they turn out to be useful, among many other tasks, in witnessing non-classicality. 

\emph{Reverse uncertainty relations - } While uncertainty relations guarantee the existence of intrinsic fluctuations in quantum theory, they usually do not let us estimate an upper bound on such fluctuations. Thus, the problem of devising reverse uncertainty relations is one of considerable theoretical interest \citep{puchala,debasis}. Below we formulate such a reverse uncertainty relation for observables.

\begin{theorem}
For arbitrary mixed state $\rho$ and observables $A$ and $B$, if the corresponding uncertainty matrix is denoted by $K = \lambda \sigma$, where $\lambda = \tr(K)$, the following  variance based reverse sum uncertainty relations hold.
\begin{equation}
\Delta A^{2} + \Delta B^{2} \leq |\langle [A,B] \rangle | + \lambda \mathcal{F}^{2}(\rho,\sigma) \leq |\langle [A,B] \rangle | + \lambda \left( 1 - S(\rho || \sigma) \right)
\label{reverse_eqn}
\end{equation}
where $\mathcal{F}$ is the fidelity between two quantum states and $S(\rho||\sigma)$ the relative entropy.
\end{theorem}
\emph{Proof-} The first inequality follows from applying the Araki-Lieb-Thirring inequality to the second term of the RHS in the uncertainty equality \eqref{urequality}. The subsequent inequality follows from the first one in two steps - first using the Fuchs van de Graaf inequality linking fidelity and trace distance, followed by Audenaert and Eisert's relation \citep{audenaert} between trace distance and relative entropy.  \qed

The problem of finding the reverse uncertainty relations in our framework turns out to be very closely related to the famous problem of determining the numerical radius of an operator in Matrix analysis. Based on the latest mathematical advances \footnote{See, for example Ref. \citep{dragomir_book}}, we derive several such reverse uncertainty relations  in the supplementary material \citep{supp}.

\emph{Reverse quantum speed limit -} Suppose the Markovian evolution of a quantum system represented by the density matrix $\rho (t)$ is given by \begin{equation}
\dot{\rho} (t) = L_t (\rho(t)),
\end{equation} We term $L_t$ as the generator of the dynamics. Suppose further that during the dynamics, the system evolves from an initial state $\rho_0$ to a final state which makes a Bures angle $\mathcal{L}$ with the initial state. The lower bound on the evolution time (equivalently the quantum speed limit) for this situation has already been explored \citep{deffner_lutz} in some detail. However, in the scenario above, we may prove \citep{supp} the following non-trivial inverse bound, i.e. upper bound on the evolution time $\tau$

\begin{equation}
\tau \leq \frac{\sin^2 \mathcal{L}_{\tau}}{\Lambda_{\text{reverse}}}
\label{pbrur_main}
\end{equation}
where $\Lambda_{\text{reverse}} = 1/ \tau \int_{0}^{\tau}  dt \left[ S(\rho_0) - \ln \tr e^{\mp L_t (\rho(t))} \right]$, and $\mathcal{L}_{\tau}$ is the Bures angle between the initial and final state. To prove this, we will follow the setting and notation laid down in Deffner and Lutz's recent work \citep{deffner_lutz}. If the initial state is $\rho_0$, the Bures angle of the state $\rho(t)$ at time $t$ with respect to the initial state is given by $\mathcal{L}= \cos^{-1} \left(\sqrt{\tr(\sqrt{\rho_0} \rho(t) \sqrt{\rho_0}} \right)$. Taking time derivative of both sides, we have the following equation \beq 2 \cos \mathcal{L} \sin \mathcal{L} |\frac{d \mathcal{L}}{dt}| = \vert \tr \left[ \rho_0 L_t (\rho(t)) \right] \vert \eeq

\noindent Now, assuming that the fidelity decreases, i.e. Bures angle $\mathcal{L}$ increases monotonically over time, as well as assuming that the quantity  $\tr \left[ \rho_0 L_t (\rho(t)) \right]$ in the RHS of the above equation is positive, we have the following equality condition

\beq \sin 2\mathcal{L} \frac{d\mathcal{L}}{dt} = \tr \left[ \rho_0 L_t (\rho(t)) \right]  \eeq

\noindent Applying the Peierls-Bogoliubov inequality \citep{bhatia} to the RHS of this equality leads to 

\begin{equation}
\sin 2 \mathcal{L} \frac{d \mathcal{L}}{dt} \geq S(\rho_0) - \ln \tr e^{-L_t (\rho(t))}
\label{pbrur}
\end{equation}

\noindent Now integrating \eqref{pbrur}, the following reverse speed limit on evolution time $\tau$ is obtained 

\begin{equation}
\tau \leq \frac{\sin^2 \mathcal{L}_{\tau}}{\Lambda_{\text{reverse}}}
\end{equation}
where $\Lambda_{\text{reverse}} = 1/ \tau \int_{0}^{\tau}  dt \left[ S(\rho_0) - \ln \tr e^{-L_t (\rho(t))} \right]$, and $\mathcal{L}_{\tau}$ is the Bures angle between the initial and final state.

If the opposite case is true, i.e.,  the quantity $\tr \left[ \rho_0 L_t (\rho(t)) \right]$ is always negative, then the reverse speed limit is given by 
\begin{equation}
\tau \leq \frac{\sin^2 \mathcal{L}_{\tau}}{\Lambda_{\text{reverse}}}
\end{equation}
where $\Lambda_{\text{reverse}} = 1/ \tau \int_{0}^{\tau}  dt \left[ S(\rho_0) - \ln \tr e^{L_t (\rho(t))} \right]$, and $\mathcal{L}_{\tau}$ is the Bures angle between the initial and final state.
\begin{figure}[h]
\includegraphics[width = 0.45 \textwidth, keepaspectratio]{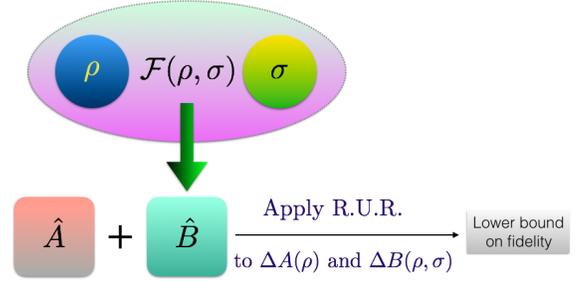}
\caption{(Color online) Diagram for the scheme to measure a lower bound to fidelity between states $\rho$ and $\sigma$.  }
\label{fid_diagram}
\end{figure}
\\

\emph{Measuring fidelity - }  Exactly measuring the fidelity between two quantum states, say $\rho$ and $\sigma$, necessitates performing complete state tomography on both of them. The computational cost for this grows exponentially with Hilbert space dimension. However, for a given state $\rho$, if we choose the observables $A$ and $B$ judiciously such that the normalized uncertainty matrix $K/ \tr(K)$ is the second state $\sigma$, then the reverse uncertainty relation \eqref{reverse_eqn} guarantees that we find a lower bound to the fidelity between $\rho$ and $\sigma$ by from simply measuring the variances  of $A$ and $B$ experimentally. The number of such measurements required does not scale exponentially with Hilbert space dimension, therefore this represents an economic way to estimate the minimum fidelity between two quantum states.

\emph{Conclusion -} We have provided several stronger uncertainty and reverse uncertainty relations, and mentioned a few of their applications in this work. However, we still think there is ample scope for discovering newer and better uncertainty relations. As an illustration, applying the methods of Ref. \citep{debasis} to our uncertainty relations would yield even tighter bounds. We hope these new uncertainty and reverse uncertainty relations can stimulate new thoughts on the so called uncertainty principle debate, as well as quantum metrology. Investigating the optimization-free uncertainty relation with the aim of relating it to the Wigner Yanase skew information may also give us new insight into the nature of uncertainty relations. These new relations have also been illustrated to result in two applications, viz., coming up with a lower bound for fidelity between two states as well as an inverse quantum speed limit for Markovian evolution. We hope that the stronger uncertainty and reverse uncertainty relations derived and discussed in this letter should empower physicists to reveal further new facts about the quantum world. In particular, we are optimistic that the reverse uncertainty relations may turn out  to be of some practical utility for metrologists, as well as be connected to some form of a reverse data-processing inequality.

\emph{Acknowledgement-} CM acknowledges Department of Atomic Energy, Governent of India, for granting of graduate research fellowship. We thank A.K. Rajagopal and L. Maccone  for discussions and comments.
\bibliographystyle{apsrev4-1} 
\bibliography{ur_ref}

\newpage

\section{Supplementary Material for ``Stronger classes of sum uncertainty and reverse uncertainty relations"}
This supplementary material is organized as follows.  The first section  deals with the possible implications of vectorization technique on  establishing another link between the resource theories of entanglement an coherence. The second section touches briefly on the uncertainty matrix and its relation with the purity of the underlying quantum system. Two subsequent sections discuss the conditions for which the stronger uncertainty relations enunciated in the paper are saturated. This is followed by a section dedicated to deriving stronger uncertainty relations for three observables starting from a generalization of the parallelogram law. Reverse uncertainty relations have been treated subsequently. We finally elaborate on the method for finding a  lower bound on fidelity, which has been mentioned in the main text.
\subsection{Vectorization}
A tool of matrix analysis useful in various applications ranging from diffusion MRI techniques \citep{mri} to finite element analysis \citep{finel} to  studying generalized Hooke's law \citep{voigt} is vectorization, which has also been used in quantum information theory in the context of quantum maps \citep{gilchrist,modi} and calculating the quantum Fisher information \citep{qfi_vec}. For any matrix $A$, this entails stacking columns of that matrix on top of one another to construct a one-dimensional array, hence called the \emph{vectorization} of the matrix $A$, which we shall denote by $|A\rangle$, and the corresponding dual vector by $\langle A|$. We mention below the following relevant relations which hold for vectorization. 

\begin{enumerate}
\item \emph{Vectorization of product of matrices: } \begin{equation}
|AB\rangle= (\mathbb{I} \otimes A) |B \rangle
\end{equation} 
\item \emph{Inner product of matrices: }\begin{equation}
\text{Tr} (A^{\dagger} B) =  \bra A|B \ket
\label{vecinnerpdt}
\end{equation}
\end{enumerate}

\subsubsection{Using vectorization to link quantum resources}
\label{append_sub_1}

The link between quantum correlations and coherence has been established in several past works \citep{uttament,mile,yao}. Let us consider a quantum state $\rho$ and ask the following question - what can the entanglement of the vectorized (and adequately normalized) state $|\rho\ket$ tell about the coherence of the original state $\rho$ ? Let us confine ourselves to the qubit case and suppose $\rho$ is a general qubit mixed state such that $\rho = \frac{1}{2} \left( \mathbb{I} + \sum \vec{r}. \vec{\sigma} \right)$. The corresponding vectorized pure two-qubit state equals = $\alpha [1+r_z,r_x + i r_y, r_x - i r_y, 1 - r_z]^T$, where $\alpha$ is a normalization constant = $1/ \sqrt{2 + r_x^2 - r_y^2 + 2r_z^2} = 1/ \sqrt{2 + 2 P^2 - C_{l_1}^{2} - 2 r_y^{2}}$, where $P = \sqrt{r_x^2 + r_y^2 + r_z^2}$ is a measure of purity of the original state, and $C_{l_1}$ is the $l_1$-norm of coherence. The concurrence of the vectorized state now reads as $E_C = 2 |\alpha|^2 |1 + r_{z}^2 - r_x^2 - r_y^2|$, which, for real qubit states  implies \beq E_c =\frac{|1+P^2 - 2  C_{l_1}^{2}|}{|1 +  P^2 - \frac{C_{l_1}^{2}}{2}|} \eeq

This is an exact equality relating three quintessentially quantum features, viz. purity content, entanglement and coherence, which holds for arbitrary real qubit states. In fact, the general expression in terms of $r_y$ also includes a fourth quantum feature, viz. imaginarity, a resource theory for which has recently been constructed \citep{imaginarity}. This may turn out to be useful when considering the conversion of one quantum resource to another.

\subsection{Uncertainty matrix and purity}
\label{append_sub_2}

Suppose a qubit state $\rho$ is of the form \beq \rho = \frac{1}{2} \left(\mathbb{I} + \vec{r}.\vec{\sigma} \right) \eeq and the operators in question are $A = \vec{\sigma}.\hat{n_1}$ and $B =\vec{\sigma}.\hat{n_2}$ respectively. Assuming that the positive sign holds in the definition of the uncertainty matrix in the main paper - the expression for the density matrix obtained through normalizing the uncertainty matrix can be shown after some tedious algebra as 

\beq 
\frac{K}{\tr{K}} = \frac{\mathbb{I}}{2} + \frac{\vec{\sigma}. \left[(\hat{n_1}\times \hat{n_2}) - (p_1\hat{n_1} + p_2\hat{n_2})\right]}{2 + p_1^2 +  p_2^2} 
\label{purity_ur}
\eeq
where $p_i = \vec{r}.\hat{n_i}$.
The length of Bloch vector $\vec{R}$ for this density matrix, which is a measure of purity, can now be shown to be equal to \beq |\vec{R}| = \frac{\sqrt{\alpha \beta - \gamma^2}}{(\alpha + \beta)/2} \eeq  where $\alpha = 1 + p_1^2$, $\beta = 1 + p_2^2$, and $\gamma = (p_1 p_2 - \hat{n_1}.\hat{n_2})$. If the state $\rho$ is highly mixed, i.e., $|\vec{r}|$ is very small, then upto leading order \beq |\vec{R}|\approx \sqrt{1 - (\hat{n_1}.\hat{n_2})^2} \label{almost_pure}\eeq
\begin{figure}[h]
\includegraphics[width = 0.35 \textwidth, keepaspectratio]{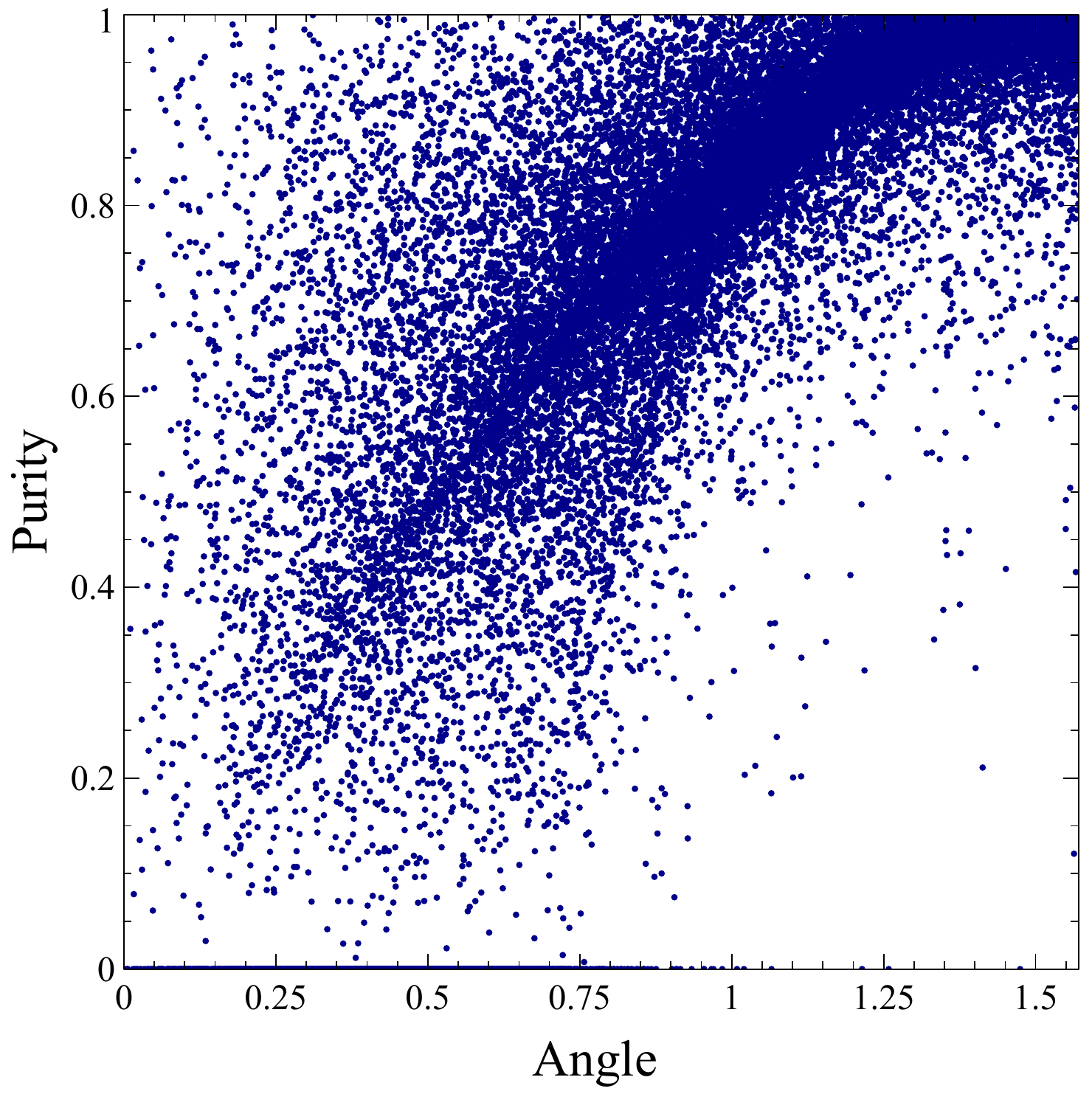}
\caption{(Color online) Plot of purity of the (normalized) uncertainty matrix vs. incompatbility captured through inner product angle between the directions $\hat{n_1}$ and $\hat{n_2}$ corresponding to observables $A$ and $B$ respectively for 60,000 iterations of random quantum states $\rho$ and random unit-vector directions $\hat{n_1}$ and $\hat{n_2}$ corresponding to observables $A$ and $B$.}
\label{purity}
\end{figure}
Clearly, the angle between the two observables $A$ and $B$ captures the incompatibility between them. If the vectors $\hat{n_1}, \hat{n_2}$ are collinear or anti-collinear, then the corresponding operators are compatible. The more the deviation from (anti)-collinearity, the more incompatible they are. From \eqref{almost_pure}, we may see in this case, that the purity of the density matrix constructed from the uncertainty matrix nicely quantifies the incompatibility of these operators. For arbitrary qubit cases, the situation is not as clear-cut. Nonetheless, Fig. \ref{purity} confirms that the general trend that, increasing purity is accompanied by increasing incompatibility between the two observables, continue to hold by and large.

\subsection{Condition for tightness of \eqref{thm1} for orthogonal states in qutrit systems}

For the Cauchy Schwarz inequality to be saturated, the corresponding vectors must be co-linear. This is not the case for the Wielandt inequality for vectors , when the saturation condition may be met even if the vectors are non-collinear, even orthogonal. If we only pick states $|\phi\ket$ as orthogonal states to $|\psi\ket$ (denoted by $|\psi^{\perp}\ket )$ in \eqref{thm1}, the condition for saturation of that uncertainty relation is given in terms of the eigenvectors of the uncertainty matrix K by the following conditions\citep{liu} together -

\begin{enumerate}
\item $|\psi\ket$ is expressible as $|\psi\ket = \frac{1}{\sqrt{2}} \left( \xi |\lambda_{\text{max}}\ket + \eta |\lambda_{\text{min}}\ket \right)$.

\item and, the perpendicular state $|\psi^{\perp}\ket$ is expressible as $|\psi^{\perp}\ket = \frac{\epsilon}{\sqrt{2}} \left( \xi |\lambda_{\text{max}}\ket - \eta |\lambda_{\text{min}}\ket \right)$. 
\end{enumerate}

Let us illustrate this for the case of a general qubit pure state and observables $A = \sigma_x$ and $B = \sigma_y$. Let us a qubit pure state $|\psi\ket = \cos \theta/2 |0\ket + e^{i \phi}\sin \theta/2  |1\ket$ in the parameter regime $\theta \in [0,\pi /4]$ \footnote{The case of $\theta \in [\pi/4 , \pi]$ can be analogusly treated.}. In this case, the uncertainty matrix $K = (C - iD) (C + iD) = C^2 + D^2 + i [C, D] $. The corresponding eigenvectors are given by 
\begin{equation} |\lambda_{\text{max}}\ket = \frac{1}{\sqrt{1 + \left( \frac{\sin \theta}{1 -\sqrt{1+ \sin ^2 \theta}}\right)^2}} \left[  1   , \frac{e^{i\phi} \sin \theta}{1 - \sqrt{1 + \sin ^2 \theta}}      \right]^{T}
\end{equation}and, \begin{equation} |\lambda_{\text{min}}\ket = \frac{1}{\sqrt{1 + \left( \frac{\sin \theta}{1 + \sqrt{1+ \sin ^2 \theta}}\right)^2}} \left[  1   , \frac{e^{i\phi} \sin \theta}{1 + \sqrt{1 + \sin ^2 \theta}}      \right]^{T}
\end{equation}.

Now if $|\psi\ket$ and $|\psi^{\perp}\ket$ can be represented in the form given above, the inequality is tight.

\subsection{Condition for tightness of \eqref{peierls} for single qubit mixed states and fixed observables}

Suppose we again choose the operators $A$ and $B$ as  $\vec{\sigma}.\hat{n_1}$ and $\vec{\sigma}.\hat{n_1}$ respectively . The uncertainty matrix $K$ can now be written as  \beq K =  \left[ 2 + p_1^2 + p_2^2 \right] \mathbb{I} + 2 \vec{\sigma}.\left[ (\hat{n_1} \times \hat{n_2}) - p_1\hat{n_1} - p_2 \hat{n_2} \right] \eeq 

We write $t=2+p_1^2+p_2^2 $ and $\vec{R}=2\left[ (\hat{n_1} \times \hat{n_2}) - p_1\hat{n_1} - p_2 \hat{n_2} \right]$, which yields the following expression for $e^{-K}$

\beq e^{-K} = t' \mathbb{I} + \vec{R'}.\vec{\sigma} \eeq

where $t' = e^{-t} \cosh |\vec{R}|$, and $\vec{R'} = - e^{-t} \frac{\sinh |\vec{R}|}{|\vec{R}|} \vec{R}$. For equality condition to be satisfied, this has to equal $\rho = \frac{1}{2} \left( \mathbb{I} + \vec{r}.\vec{\sigma} \right)$. That is, the following sets of non-linear equations have to be simultaneously satisfied -

\begin{eqnarray}
 e^{-t} \cosh |\vec{R}| = \frac{1}{2} \\
e^{-t} \frac{\sinh |\vec{R}|}{|\vec{R}|} \vec{R} = -\frac{\vec{r}}{2}
\end{eqnarray}

If these set of equations do not have a solution - this inequality is not saturated. 

\subsection{Uncertainty relation for three incompatible observables}
\begin{figure}
\includegraphics[width = 0.45 \textwidth, keepaspectratio]{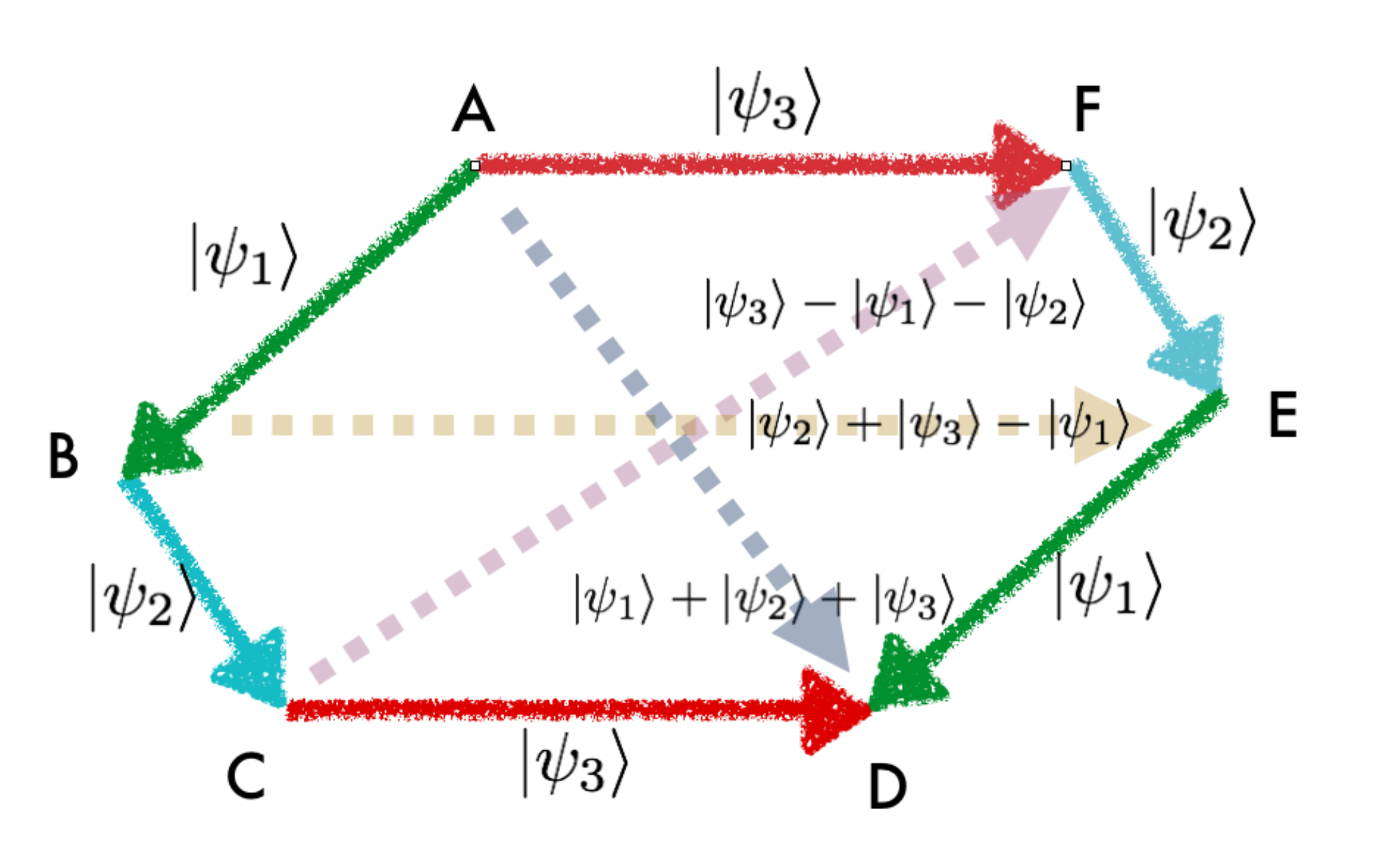}
\caption{(Color online) One generalization of the parallelogram law for parallelo-hexagons.}
\label{hexagon}
\end{figure}

Let us confine ourselves to three observables $A, B, C$. Let us denote $P = A - \bra A\ket , Q = B - \bra B\ket, R = C - \bra C\ket$. Now, let us consider the parallelo-hexagon, i.e. the hexagon ABCDEF whose opposite pairs are parallel and equal in length, in Fig. \ref{hexagon}. Let us choose vectors $\vec{AB} = |\psi_1\ket = (P  + ik_1 Q) |\psi\ket, \vec{BC} = |\psi_2\ket = (Q + i k_2 R) |\psi\ket, \vec{CD} = |\psi_3 \ket = (P +  i k_3 R) |\psi\ket$. The numbers $k_1,k_2,k_3 \in \lbrace -1, +1 \rbrace$,  are chosen based on whether the corresponding commutators give rise to positive or negative real numbers, in a similar way to Ref. \citep{mpur}. Now, the following equality holds \cite{douglas} for the parallelo-hexagon. 

\begin{figure}
\includegraphics[width = 0.45 \textwidth, keepaspectratio]{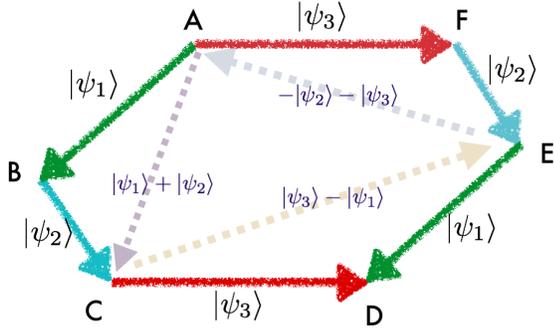}
\caption{(Color online) Another generalization of the parallelogram law for parallelo-hexagons.}
\label{hexagon2}
\end{figure}
\begin{equation}
||\vec{AB}||^2 + ||\vec{BC}||^2 + ||\vec{CD}||^2 = \frac{1}{4} \left( ||\vec{AD}||^2 + ||\vec{BE}||^2 + ||\vec{CF}||^2 \right)
\end{equation}

The LHS quantity of this equality can now be shown to equal $\Delta A^2 + \Delta B^2 + \Delta C^2 - |\langle [A,B] \rangle|- |\langle [B,C] \rangle| - |\langle [A,C] \rangle| =\frac{1}{2} \left[ \Delta A^2 + \Delta B^2 - |\langle[A,B] \rangle| \right] + \frac{1}{2} \left[ \Delta A^2 + \Delta C^2 - |\langle[A,C] \rangle| \right] + \frac{1}{2} \left[ \Delta B^2 + \Delta C^2 - |\langle[B,C] \rangle| \right]$. That is, broken down into sums of two-observable Robertson uncertainty relations. Thus, since the RHS is always positive semi-definite, applying the generalized Wielandt inequality to the norms of each of the vector $\vec{AD}, \vec{BE}, \vec{CF}$, coupled with any qutrit pure state vectors, give rise to stronger uncertainty relations of the type similar to \eqref{thm1}.

Is this the only form of uncertainty relation derivable ? It turns out, there exists another \citep{douglas} geometric identity (please see Fig. \ref{hexagon2}).  

\begin{equation}
||\vec{AB}||^2 + ||\vec{BC}||^2 + ||\vec{CD}||^2 = \frac{1}{3} \left( ||\vec{AC}||^2 + ||\vec{CE}||^2 + ||\vec{EA}||^2 \right)
\end{equation}

Proceeding similarly as before, it is possible to obtain another set of uncertainty relations for observables $A$, $B$, $C$ from this equality too.


\subsection{Further reverse uncertainty relations}

Let us first recall the definition of the \emph{numerical radius} \citep{bhatia} of an operator.

\textbf{Definition (Numerical Radius) :} \emph{The numerical radius of an operator T is the number}
\begin{equation}
w(T) = \sup_{|||\phi \ket|| =1} |\bra \phi |A| \phi \ket |
\label{n_rad}
\end{equation} 

Thus, it is clear that for pure states $|\psi\ket$, the quantity which crucially appears at the right hand side of for every stronger uncertainty relation, i.e., $|| C \mp iD |\psi\ket ||^2$ is upper bounded by nothing but the square of the numerical radius of the uncertainty matrix $K$. Thus, the following reverse uncertainty relation holds
\begin{equation}
\Delta A^2 + \Delta B^2 \leq |\bra [A,B]\ket| + w^2(K)
\label{rur_gen}
\end{equation}
It follows that any upper bound to the numerical radius of $K$ automatically furnishes a reverse uncertainty relation. In the following, we summarize the underlying mathematical inequalities for numerical radius of an arbitrary operator $T$ and then write down the corresponding reverse uncertainty relation thus obtainable.

\noindent \textbf{Berger inequality (Berger \citep{berger}, 1965)} \emph{For any natural number $n$ and any operator $T$, \beq w(T^n)  \leq w^n (T) \eeq }

The corresponding reverse uncertainty relation reads as 

\begin{equation}
\Delta A^2 + \Delta B^2 \leq |\bra [A,B]\ket| + w^4(\sqrt{K})
\end{equation}
Now, it is relatively easy to note that the numerical radius is trivially upper bounded by the operator norm of the positive semi-definite operator $T$. Perhaps more non-trivial is the following inequality.
\\
\noindent\textbf{Kittaneh inequality (Kittaneh \citep{kittaneh}, 2003)} \emph{For  any operator $T$, \beq w(T^n)  \leq \frac{1}{2} \left( || T || +||T^2||^{1/2}\right) \eeq }

The reverse uncertainty relation corresponding to this inequality reads as
\begin{equation}
\Delta A^2 + \Delta B^2 \leq |\bra [A,B]\ket| + \frac{1}{4}\left| ||\sqrt{K}|| + \sqrt{||K||} \right|^2
\end{equation}

A further generalization of Kittaneh's original inequality for numerical radius comes via the following inequality -
\\
\noindent \textbf{El Haddad and Kittaneh inequality (El Haddad and Kittaneh \citep{kittaneh_elhaddad}, 2007)} \emph{For  any operator $T$ and the  adjoint operator $T^*$, if we denote $|T| = (T^*T)^{1/2}$, then $\forall \alpha \in (0,1), r \geq 1$ \beq w^{2r} (T)  \leq  || \alpha |T|^{2r} + (1-\alpha) |T^*|^{2r}  || \eeq }
The above result may also be used to yield reverse uncertainty relations.


\subsection{Measuring Fidelity}

Let us concern ourselves with the problem - how does one measure the fidelity between two qubit states $\rho$ and $\sigma$ ? Suppose we are given two states $\rho$ and $\sigma$, as well as the apparatus to measure one observable $A$. We employ the trick of introducing another observable $B$ dependent on both the states $\rho$ and $\sigma$, such that $\sigma$ turns out to be (normalized) uncertainty matrix corresponding to observables $A$, $B$ and the state $\rho$. The next step is simply to find out the sum of variances of $A$ and $B$ - which automatically furnishes a lower bound on the fidelity between the original pair of states $\rho$ and $\sigma$ vide \eqref{reverse_eqn}.

Suppose $\rho = \frac{1}{2} \left[ \mathbb{I} + \vec{r}. \vec{\sigma} \right]$, and $\sigma = \frac{1}{2} \left[ \mathbb{I} + \vec{s}. \vec{\sigma} \right] $. Suppose further without loss of generality that the experimental setup allows one to measure the observable $A =\sigma.\hat{m}$. The goal is to build such an observable $B = \lambda \vec{\sigma}.\hat{n}$, where $\lambda$ is a scale factor and $\hat{n}$ an unit vector, such that the corresponding normalized uncertainty matrix $K/\tr K$ equals $\sigma$, i.e.,
\begin{equation}
 \frac{\mathbb{I} }{2} +  \frac{1}{2}  \vec{s}. \vec{\sigma}  = \frac{\mathbb{I}}{2} + \frac{\vec{\sigma}. \left[\lambda (\hat{m}\times \hat{n}) - (p_1\hat{m} +\lambda^2  p_2\hat{n})\right]}{2 + p_1^2 + \lambda^2 p_2^2} 
\end{equation}

Thus, comparing the vectors, one reaches the following vector equation \beq \vec{s} =  \frac{\lambda (\hat{m}\times \hat{n}) - (p_1\hat{m} +\lambda^2  p_2\hat{n})}{1 + \frac{1}{2} p_1^2 + \frac{\lambda^2}{2} p_2^2}  \eeq

Now our goal is to solve these set of linear equations to obtain the unit vector $\hat{n}$ with adequate scaling $\lambda$. By choosing judiciously the value of $\lambda$. Since this is a set of linear equations - this can always be done. 

\end{document}